%% file: main.tex
\documentclass[10pt]{article}
\input{hicss-packages.tex}

\newcommand{\rev}[1]{\textcolor{black}{#1}}

\title{Mobile App Security Trends and Topics: \\An Examination of Questions From Stack Overflow}

\author{Timothy Huo \\
 University of Hawaiʻi at Mānoa \\
 {\underline{\href{mailto:thuo@hawaii.edu}{thuo@hawaii.edu}} }\\ \\
 Anthony Peruma \\
 University of Hawaiʻi at Mānoa \\
 {\underline{\href{mailto:peruma@hawaii.edu}{peruma@hawaii.edu}} } \\ \And
 Ana Catarina Araújo \\
 University of Hawaiʻi at Mānoa \\
 {\underline{\href{mailto:acoa@hawaii.edu}{acoa@hawaii.edu}} } \\ \\
 Rick Kazman\\
 University of Hawaiʻi at Mānoa \\
 {\underline{\href{mailto:kazman@hawaii.edu}{kazman@hawaii.edu}} } \\ \And
 Jake Imanaka \\
 University of Hawaiʻi at Mānoa \\
 {\underline{\href{mailto:jimanaka@hawaii.edu}{jimanaka@hawaii.edu}} } \\ \\
 }

\date{}

\newcommand{\RQA}{\textbf{RQ1}: How have mobile security discussions on Stack Overflow grown over the years?}
\newcommand{\RQAA}{\textbf{RQ$_{1.1}$}: How have mobile security questions expanded over the years?}
\newcommand{\RQAB}{\textbf{RQ$_{1.2}$}: What tags are utilized for mobile security questions?}

\newcommand{\RQB}{\textbf{RQ2}: What security challenges do mobile developers face?}

\begin{document}
\maketitle
\begin{abstract}
The widespread use of smartphones and tablets has made society heavily reliant on mobile applications (apps) for accessing various resources and services. These apps often handle sensitive personal, financial, and health data, making app security a critical concern for developers. While there is extensive research on software security topics like malware and vulnerabilities, less is known about the practical security challenges mobile app developers face and the guidance they seek. \rev{In this study, we mine Stack Overflow for questions on mobile app security, which we analyze using quantitative and qualitative techniques.} 
The findings reveal that Stack Overflow is a major resource for developers seeking help with mobile app security, especially for Android apps, and identifies seven main categories of security questions: Secured Communications, Database, App Distribution Service, Encryption, Permissions, File-Specific, and General Security. Insights from this research can inform the development of tools, techniques, and resources by the research and vendor community to better support developers in securing their mobile apps.
\end{abstract}

\subsubsection*{Keywords:}

Stack Overflow, Android, iOS, Security

\section{Introduction}
\label{sec:introduction}
Technology advancements, such as integrated development environments (e.g., Xcode, Android Studio, etc.), software frameworks/libraries (e.g., OkHTTP, Lottie, etc.), and platforms (e.g., Firebase, Supabase, etc.) have made it possible for almost anyone to create mobile applications (apps) without extensive programming knowledge (\cite{allen2021android,sahar2021ios}). This has led to a significant increase in the volume of apps available on app distribution service providers like the Google Play Store and Apple App Store, ranging from simple timezone conversion apps to more sophisticated apps for online financial transactions and health and fitness tracking (\cite{AppStore_stats}). However, this ease of app development also leads to the development of poorly designed and implemented apps. This, in turn, can negatively impact the user experience (\cite{Khalid2015Reviews}) and even result in end-users installing unsafe apps that expose sensitive data or act as a vector for malware (\cite{qamar2019mobile}). While there are best practices, tools, and techniques to help developers and project teams construct high-quality, secure software systems (\cite{pressman2019software}), one should remember that mobile apps, even though technically software systems, have different characteristics from non-mobile systems. For instance, unlike non-mobile systems, mobile apps need to be energy and resource conscious (\cite{CaretteSANER2017}), have access to fewer default permissions due to tighter sandboxing (\cite{Scoccia2019SCAM}), are vetted through centralized app stores (\cite{LinWWW2021}), rely more on inconsistent and untrusted network connectivity (\cite{greenwood2014smv}), and depend more heavily on external libraries for added functionality (\cite{feal2021don}), among other characteristics (\cite{Aldayel2017AppDevelopment,Flora2014MobileApps}). 
Hence, it is essential to study mobile apps separately from other software systems to address their unique challenges. Therefore, in this study, we mine the questions developers ask about mobile app security on Stack Overflow-- one of the largest online programming-specific question-answer sites, with over 20 million questions, answers, and users (\cite{SO_QuestionsUsers}).

\subsection{Motivating Example}
Consider the question the developer asks in Quote \ref{Quote:quote-motivation}. This type of security question is specific to mobile apps-- it deals with challenges using the device's fingerprint sensor on different devices (i.e., app portability). This security issue is not usually seen in non-mobile systems, where in order to increase the adoption of their app, the developer has to support multiple devices while not compromising on security. However, since smartphones are manufactured by multiple vendors, with these vendors possibly using different variants of hardware components and software libraries, app portability becomes challenging. Hence, while general research on software security is essential, we also need to understand the intricacies of mobile app security, especially the real-world challenges developers face in securing their apps while supporting a large and diverse user base.

\begin{center}
\fbox{\parbox{\dimexpr\linewidth-2\fboxsep-2\fboxrule\relax}{\centering
\textbf{SecurityException while checking if fingerprints are enrolled in Samsung Phones}
\begin{flushleft}
\textit{``I am using a lockscreen with fingerprint in my app. While it works seamlessly with other phones having fingerprint sensor, samsung users are facing some SecurityException as I can see in my google console reports. I am having a hard time figuring it out as I have no samsung phones to test. Till now it has happened in Galaxy J7 and Grand Prime Plus"}
\end{flushleft}
}}
\captionof{Quote}{An example of a developer asking the community for help in solving an issue in securing their mobile app (\cite{quote-motivation}).\label{Quote:quote-motivation}} 
\end{center}

\subsection{Goal \& Research Questions}
\rev{Grounded in theories of socio-technical systems and knowledge sharing in software development communities,} the goal of this study is to \textit{understand the key trends, topics, and challenges around developer discussions on securing their mobile apps}. We envision our findings helping researchers and educators better understand these trends and challenges in order to improve security training materials and tools tailored to the needs of app developers. Moreover, the results can aid developers in better anticipating and addressing potential vulnerabilities in their mobile applications by planning for security issues that commonly arise in practice. Below are the research questions (RQs) for this study along with the reasons for their inclusion: 

\noindent\textbf{\RQA} This research question aims to understand the extent to which developers turn to the community for help with securing their mobile apps. Through this research question, we measure the yearly growth of mobile security questions on Stack Overflow and examine the types and growth of tags developers associate with these questions. The findings from this research question form a starting point for understanding the trends and challenges of mobile app security in a real-world setting.

\noindent\textbf{\RQB} Through this research question, we aim to discover the specific challenges mobile app developer face when securing their apps. Hence, we utilize natural language processing techniques to present a granular view of the challenging topics in mobile app security.

\section{Related Work} 
\label{sec:related}
Previous studies have investigated mobile app security discussions on Stack Overflow. \textcite{Tahaei2022EuroUSEC} analyzed privacy and permission-related posts for Android and iOS health apps, while \textcite{beyer2014manual} manually examined Android-related posts, finding that security questions were less common than other topics. \textcite{Fischer2017SP} classified security-related code snippets in Android posts and analyzed their presence in Android apps. Studies have also investigated general security-related discussions on Stack Overflow. \textcite{Yang2016security} highlighted mobile security as a topic, while \textcite{Lopez2018SEAD} performed thematic analyses on security questions and comments. \textcite{TahaeiCHI2020} conducted topic modeling on privacy-related questions, and \textcite{Licorish2021QRSC} examined security faults in code snippets. \textcite{bayati2016information} quantitatively analyzed information security questions, and \textcite{Hong2021ACSAC} and \textcite{Chen2020ASONAM} developed techniques to detect insecure code snippets in Stack Overflow. \textcite{Rosen2016MobileDevelopers} and \textcite{Linares2014MSR} identified common topics, but security was not a primary focus. \rev{Similarly, \textcite{BEDDIAR2020} analyzes and classifies Android API-related discussions on Stack Overflow,  but does not address API security issues or topics. \textcite{Ahmad2019} focuses on analyzing non-functional requirements (NFRs) for iOS app development based on Stack Overflow posts, but security is not mentioned as one of the NFRs analyzed in this study.} \rev{A study conducted by \textcite{Li2021} analyzed discussions among Android developers on Reddit. The study found that privacy concerns were not often discussed in relation to specific app design contexts. However, these concerns were actively debated when prompted by external privacy-enhancing restrictions.}

The research community has found that mining Stack Overflow is valuable for studying real-world software engineering practices, including mobile app development and security. Our work extends knowledge of app development challenges by analyzing a large, recent dataset of mobile app security discussions.

\section{Method}
\label{sec:method}
In this section, we provide details about the methodology for our study. Furthermore, the artifacts from this study are available for download \footnote{\url{https://zenodo.org/doi/10.5281/zenodo.13753125}}. 

\subsection{Dataset}
\label{Section:method_dataset}
To obtain the Stack Overflow posts for this study, we used the Stack Exchange Data Explorer\footnote{https://data.stackexchange.com/}. We ran three queries to collect mobile security posts: one for all questions and two for accepted and non-accepted answers. We extracted posts from the inception of Stack Overflow (September 2008) to December 31, 2022. Similar to prior research (\cite{Peruma2021StackOverflow}), our search process involves querying the tag and title of questions, as developers concisely state the main objective of the question in the title (\cite{Rosen2016MobileDevelopers}). We refrained from searching keywords in the question body as these may not necessarily indicate relevance to mobile security. 

\noindent\textbf{Tag Query:}
The tag search retrieves questions with tags containing `\%security\%' plus `\%\textless android\%' or `\%\textless ios\%'. The percentage sign (\%) surrounding the tag indicates a wildcard search. Further, to exclude false positives, the less-than sign (\textless) indicates that the tag must start with the word `android' or `ios'. For example, when searching questions for Android and iOS without the less-than sign, questions with tags such as `kiosk’ appear because they have the characters `ios' in the word; these questions may not be completely associated with mobile security. On the other hand, by keeping the tag search open at the end of the word, such as `\%\textless android\%', the query can capture tags such as `android-security'. Furthermore, a query constructed using only `\%security\%' and `\%\textless android\textgreater \%' would exclude questions with the `android-security' tag, which has over 600 questions. 

\noindent\textbf{Title Query:}
The condition for querying titles is similar to the tag query. Our query retrieves questions where the title contains the word `\% security \%' plus `\% android \%' or `\% ios \%'. Note: We surround each query term with a white space to help exclude any false positives, similar to the less-than sign for the tag query.

We limited our query to Android and iOS, as these are the leading mobile operating systems (\cite{MobileOS_market01}). Furthermore, developers often include these terms when seeking help with a specific framework or library, which can be security-related.

\subsection{RQ Analysis}
Similar to prior research (\cite{Peruma2021StackOverflow}), we follow a mixed-method approach consisting of quantitative and qualitative investigations to answer our RQs. More specifically, our quantitative analysis involves using well-established statistical approaches and performing a topic modeling analysis. In our qualitative analysis, we manually examine a statistically significant sample of the data.  
In Section \ref{sec:results}, \rev{we elaborate on the specific techniques we utilize to answer each RQ.}

\section{Results}
\label{sec:results}
This section reports on the findings of our study by answering our RQs. 

\subsection{\RQA}
\noindent\textbf{RQ Motivation:}
This RQ provides insight into the extent to which mobile developers turn to the community for help with security challenges for their apps and how frequently they receive assistance. Furthermore, by analyzing the tags developers use, we gain a high-level understanding of the types and trends of common technologies and concepts developers find challenging. 

Hence, this RQ is composed of two sub-RQs that examine the volume and growth of mobile security posts on Stack Overflow over time by analyzing specific attributes of the posts. RQ$_{1.1}$ investigates the volume of mobile security questions and their answers. RQ$_{1.2}$ investigates the type and growth of the common tags developers use for mobile security questions.

\vspace{1mm}
\noindent\textit{\RQAA} \newline
Executing the queries described in Section \ref{Section:method_dataset} yields 5,759 questions related to mobile security, with a majority (i.e., 4,586 or 79.63\%) of these questions receiving an answer. Looking at the answered questions, we observe that there are 2,387 (or 41.45\%) questions with accepted answers and 3,372 (or 58.55\%) questions that do not have an accepted answer. Finally, only 1,173 (or 20.37\%) questions did not receive an answer.

\begin{table}
\centering
\caption{Mobile security questions asked each year.}
\label{Table:questions_year}
\resizebox{\columnwidth}{!}{%
\begin{tabular}{@{}rrrrr@{}}
\toprule
\multicolumn{1}{c}{\multirow{2}{*}{\textbf{Year}}} &
  \multicolumn{1}{c}{\multirow{2}{*}{\textbf{\begin{tabular}[c]{@{}c@{}}Total\\ Questions\end{tabular}}}} &
  \multicolumn{3}{c}{\textbf{Questions With}} \\ \cmidrule(l){3-5} 
\multicolumn{1}{c}{} &
  \multicolumn{1}{c}{} &
  \multicolumn{1}{c}{\textbf{\begin{tabular}[c]{@{}c@{}}No \\ Answers\end{tabular}}} &
  \multicolumn{1}{c}{\textbf{\begin{tabular}[c]{@{}c@{}}Non-Accepted \\ Answers\end{tabular}}} &
  \multicolumn{1}{c}{\textbf{\begin{tabular}[c]{@{}c@{}}Accepted\\ Answers\end{tabular}}} \\ \midrule
2022 & 253 & 118 & 83  & 52  \\
2021 & 272 & 94  & 101 & 77  \\
2020 & 414 & 119 & 164 & 131 \\
2019 & 375 & 106 & 147 & 122 \\
2018 & 455 & 99  & 191 & 165 \\
2017 & 647 & 149 & 267 & 231 \\
2016 & 872 & 195 & 336 & 341 \\
2015 & 631 & 108 & 267 & 256 \\
2014 & 510 & 87  & 200 & 223 \\
2013 & 473 & 58  & 184 & 231 \\
2012 & 464 & 31  & 146 & 287 \\
2011 & 304 & 8   & 93  & 203 \\
2010 & 83  & 1   & 19  & 63  \\
2009 & 6   & 0   & 1   & 5   \\ \bottomrule
\end{tabular}%
}
\end{table}

Next, we examine the yearly breakdown of questions by examining the date when the question was posted. As shown in Table \ref{Table:questions_year}, 2016 witnessed the highest amount of mobile security questions (872, to be precise) posted by developers, followed by 2017 and 2015. Furthermore, the median number of yearly mobile security questions is 434.50, while the median number of answers in a year is 535. Moreover, we also observe that for each year, more questions receive an answer than questions that do not receive an answer. 

Moving on, to understand the reason for the increase in questions in 2016, we reviewed the Title of the questions posted in 2015, 2016, and 2017. Our analysis shows a rise in questions around the Google Play Store, such as help resolving \texttt{X509TrustManager}\footnote{\url{https://developer.android.com/reference/javax/net/ssl/X509TrustManager}} warnings. This aligns with Google Play blocking apps containing unsafe implementations \texttt{X509TrustManager} in May 2016 (\cite{GooglePlay-x509trustmanager}). Additionally, iOS developers needed help with \texttt{App Transport Security}\footnote{\url{https://developer.apple.com/documentation/bundleresources/information\_property\_list/nsapptransportsecurity}} issues on iOS 9 or 10, which were released during this period and could explain the rise in questions on this topic. 

Finally, we examine the median time between a developer asking a question and receiving a response. We observe that it takes approximately 87.1 minutes (or 1.45 hours) for a mobile security question to receive its first answer. In contrast, in a study on refactoring questions on Stack Overflow, the authors observe a median time interval of 0.27 hours (\cite{Peruma2021StackOverflow}). This comparison shows that answering mobile security questions requires developers with specialized knowledge, hence the relatively long wait for a response.

\vspace{1mm}
\noindent\textit{\RQAB}\newline
This sub-RQ looks at the tags developers use to label mobile security questions. Tags help developers categorize and locate questions that are relevant to them. Stack Overflow maintains a predefined list of tags from which developers can select one or more to assign to a question. Our extracted dataset contains a total of 1,793 unique tags. From this set, we derive the number of times each tag occurs in the dataset. Table \ref{Table:TagCount} shows the top 5 frequently occurring tags, with `android' claiming first place with 4,022 (or 17.72\%) instances, followed by `security' with 3,612 (or 15.92\%) instances, and `ios' with 1,609 (or 7.09\%) instances. 

\begin{table}
\centering
\caption{Top five occurring tags for mobile security questions.}
\vspace{-2mm}
\label{Table:TagCount}
\begin{tabular}{@{}lrr@{}}
\toprule
\multicolumn{1}{c}{\textbf{Tag Name}} & \multicolumn{1}{c}{\textbf{Occurrence}} & \multicolumn{1}{c}{\textbf{Percentage}} \\ \midrule
android          & 4,022           & 17.72\%          \\
security         & 3,612           & 15.92\%          \\
ios              & 1,609           & 7.09\%           \\
java             & 660             & 2.91\%           \\
android-security & 623             & 2.75\%           \\
\textit{others}  & \textit{12,168} & \textit{53.62\%} \\
\textbf{Total}   & \textbf{22,694} & \textbf{100\%}   \\ \bottomrule
\end{tabular}
\end{table}

Even though developers utilize various tags for their mobile security questions, these tags can be grouped into categories. Similar to (\cite{Peruma2021StackOverflow}), we manually examined a statistically significant sample of frequently utilized tags and grouped related tags into categories. To this extent, three authors analyzed 413 distinct tags, which equates to a confidence level of 99\% and a confidence interval of 5\%. Each author independently evaluated and grouped related tags into specific categories, resolving any disagreements through discussions and agreeing on the finalized set of categories. As shown in Table \ref{Table:TagCategory}, our analysis yields 10 categories. Most of the tags are part of the \textit{Security Concepts/Features} category, which includes general security concepts like `oauth' and more mobile-specific features such as `android-securityexception'. The next highest category is \textit{Framework/Library/API}, which includes mobile and non-mobile technologies developers utilize in building apps, such as `cordova' and `angularjs'. Additionally, developers also use general \textit{Programming/Software Engineering Concepts} tags such as `debugging' and `web-services'. We also observe developers using tags related to mobile and non-mobile app development \textit{Tools} such as `xcode' and `eclipse'. Similarly, developers also use \textit{Operating System} tags like `android' and `macos'. The \textit{General Technology Concept} category includes tags like `browser' and `screenshot', while the \textit{Programming Language} category includes languages like `java' and `swift'. Developers also refer to \textit{Hardware}, which includes mobile devices, such as `ipad' and `samsung-mobile'. Finally, in the last two categories, \textit{Storage} includes database technologies like `sqlite', while \textit{General Mobile App Functionality} includes features specific to mobile apps like `in-app-billing' and `push-notification'.

\begin{table}
\centering
\caption{Breakdown of the volume of instances for each tag category.}
\label{Table:TagCategory}
\resizebox{\columnwidth}{!}{%
\begin{tabular}{@{}lrr@{}}
\toprule
\multicolumn{1}{c}{\textbf{Tag Category}} & \multicolumn{1}{c}{\textbf{Count}} & \multicolumn{1}{c}{\textbf{Percentage}} \\ \midrule
Security Concepts/Features                & 154 & 37.29\% \\
Framework/Library/API                     & 101 & 24.46\% \\
Programming/Software Engineering Concepts & 42  & 10.17\% \\
Tools                                     & 29  & 7.02\%  \\
Operating System                          & 25  & 6.05\%  \\
General Technology Concept                & 19  & 4.60\%  \\
Programming Language                      & 12  & 2.91\%  \\
Hardware                                  & 12  & 2.91\%  \\
Storage                                  & 10  & 2.42\%  \\
General Mobile App Functionality          & 9   & 2.18\%  \\ \bottomrule
\end{tabular}%
}
\end{table}

\begin{tcolorbox}[top=0.5pt,bottom=0.5pt,left=1pt,right=1pt]
\textbf{Summary for RQ1.}
Stack Overflow remains a well-known platform for developers seeking help with mobile security issues. Most of the questions are about the security of Android applications and they are tagged with various security-related labels. Furthermore, significant changes made by Google and Apple to the mobile ecosystem may cause a sudden surge in the number of questions.
\end{tcolorbox}

\subsection{\RQB}
\begin{table*}
\centering
\caption{Volume of questions for each LDA topic and reviewed by the authors, and a partial set of associated words.}
\label{Table:LDA}
\resizebox{\textwidth}{!}{%
\begin{tabular}{@{}lrrrl@{}}
\toprule
\multicolumn{1}{c}{\multirow{2}{*}{\textbf{Topic}}} &
  \multicolumn{3}{c}{\textbf{Questions}} &
  \multicolumn{1}{c}{\multirow{2}{*}{\textbf{Topic Representative Words}}} \\
\multicolumn{1}{c}{} &
  \multicolumn{1}{c}{\textbf{Count}} &
  \multicolumn{1}{c}{\textbf{Percentage}} &
  \multicolumn{1}{c}{\textbf{Manually Reviewed}} &
  \multicolumn{1}{c}{} \\ \midrule
General Security &
  3,100 &
  53.83\% &
  342 &
  \begin{tabular}[c]{@{}l@{}}application, android, ios, implement, login, credentials,\\password, server, device, secure, api, token, client, access, user\end{tabular} \\
Secured Communications &
  843 &
  14.64\% &
  265 &
  \begin{tabular}[c]{@{}l@{}}certificate, error, ssl, http, transport, load, ats, https, network,\\ domain, tls, cert, connect, url, webview\end{tabular} \\
Databases &
  668 &
  11.60\% &
  245 &
  \begin{tabular}[c]{@{}l@{}}database, firebase, rule, create, read, write, delete, firestore,\\db, query, data, add, update, id, uid, field\end{tabular} \\
App Distribution Service &
  386 &
  6.70\% &
  193 &
  \begin{tabular}[c]{@{}l@{}}google, play, version, apk, apps, vulnerability, warning, console,\\cordova, developer, store, error, policy, email, package, rejected\end{tabular} \\
Encryption &
  329 &
  5.71\% &
  178 &
  \begin{tabular}[c]{@{}l@{}}keys, keystore, encryption, private, string, public, encrypt,\\encrypted, algorithm, secret, keychain, decrypt, rsa, cipher, aes\end{tabular} \\
Permissions &
  253 &
  4.39\% &
  153 &
  \begin{tabular}[c]{@{}l@{}}permission, granted, denied, request, manifest, infoplist,\\securityexception, exception, intent, broadcast, sms\end{tabular} \\
File Specific &
  180 &
  3.13\% &
  123 &
  \begin{tabular}[c]{@{}l@{}}project, build, gradle, run, config, path, uri, file, folder,\\directory, image, videos, studio, ipa, apk\end{tabular} \\
Total &
  \textit{5,759} &
  \textit{100.00\%} &
  \textit{1,499} &
  \textit{--} \\ \bottomrule
\end{tabular}%
}
\end{table*}
\noindent\textbf{RQ Motivation:}
The prior RQ shows Stack Overflow is a popular venue for asking mobile security questions and the types of tags developers utilize to categorize their questions. In this RQ, we utilize natural language processing techniques to better understand the security challenges faced by mobile app developers. To do this, we analyze the actual content of the questions asked. By focusing on the body of the questions, which is written in natural language, we can gain more insight into the difficulties that developers encounter when securing their apps, as compared to just looking at the tags used.

Prior to analyzing the question body, we normalize the text (i.e., we do text preprocessing). Our activities include converting all characters to lowercase, removing code snippets, formatting tags, digits, and special characters, and expanding contractions. Additionally, we also remove stopwords, both standard and custom words. Next, we perform a topic modeling analysis on the body of the question using the Latent Dirichlet Allocation (LDA) algorithm (\cite{Blei2003LDA}). This type of analysis is frequently utilized in similar studies (e.g., \cite{ Rosen2016MobileDevelopers,Peruma2021StackOverflow}). Our approach involves running multiple execution cycles of the LDA analysis. We start with two topics and then increase the number of topics by one in each subsequent cycle until we have 50 topics, with each cycle having 150 passes and iterations. We determine the optimal model by calculating the topic coherence (\cite{Roder2015Coherence}) and manually inspecting the output (\cite{Sievert2014LDAvis}) of each model. Our analysis determined that the most optimum model is the seven-topic model. Since the LDA algorithm does not provide a name for the generated topics, the authors manually examined the distribution of words across the topics to determine the topic names. Additionally, to understand the distribution of questions among these topics, we assigned each question to its dominant topic. Furthermore, to help us contextualize the topic assignments, we reviewed 1,499 questions, a stratified statistically significant (95\% confidence level and 5\% confidence interval) sample of questions (in each topic). Table \ref{Table:LDA} shows a breakdown of the seven topics and the words typically associated with each topic. Questions on ``General Security'' practices and concerns occur the most, followed by more specialized topics like ``Secured Communications'', ``Database'', and so on. Based on our manual review, the subsections below describe each topic and include a suitable example for each topic.

\noindent\textit{\textbf{Secured Communications.}} While mobile apps provide users with the flexibility of accessing various online services from their smartphone/tablet, it is essential that these apps provide the user with a secure environment to access these services. Mobile apps should secure the data stored locally on the device and securely communicate it to and from the device while preserving its integrity. Apps can provide a secure connection using technologies such as SSL, TLS, and HTTPS that utilize digital certificates for identity verification. 

\noindent Our examination of these questions shows that both Android and iOS app developers need help in this area. A common pain point for iOS app developers is with \texttt{App Transport Security}, such as either including or excluding specific domains. We also see app developers struggling with certificates, such as generating and accepting self-signed certificates. For example, in Quote \ref{Quote:rq2-communication}, an iOS app developer faces an issue with their app accepting a self-signed server certificate. Additionally, we see questions about making and debugging SSL connections.

\begin{center}
\fbox{\parbox{\dimexpr\linewidth-2\fboxsep-2\fboxrule\relax}{\centering
\textbf{How to use NSURLConnection to connect with SSL for an untrusted cert?}
\begin{flushleft}
\textit{``I have a simple code to connect to a SSL webpage, however, it gives an: untrusted server certificate. Is there a way to set it to accept connections anyway (just like in a browser you can press accept) or a way to bypass it?"}
\end{flushleft}
}}
\captionof{Quote}{An example of an iOS secured communications question for assistance with NSURLConnection and SSL
(\cite{rq2-communication}).\label{Quote:rq2-communication}}
\end{center}

\noindent\textit{\textbf{Database.}} A database forms an integral part of many mobile apps that require a mechanism to store and retrieve data, usually for offline functionality or caching to improve performance. This data can be in the form of user information and app data. While storing data is essential, securing it is equally important. Unauthorized access to user data can result in serious consequences such as theft of personal information, identity theft, or other fraudulent activities.

\noindent Our analysis of questions associated with this topic shows access control as a common area of concern for app developers. This includes issues related to database authentication and permissions to perform CRUD operations. Furthermore, developers face most of these security pain points when utilizing Firebase/Firestore. For example, in Quote \ref{Quote:rq2-database}, an app developer seeks help with providing an authenticated user read/write access to a Firebase database.

\begin{center}
\fbox{\parbox{\dimexpr\linewidth-2\fboxsep-2\fboxrule\relax}{\centering
\textbf{Firebase error: Permission denied. Unable to read/write from Firebase Database}
\begin{flushleft}
\textit{``I need to give read/write access only to authenticated users, but it seems like Firebase is not recognizing that the user is authenticated. After I sign in the user with email and password, I am assuming user is authenticated and should be allowed to read/write from database. However permission is denied. Please help, how to authenticate a registered user to access database in Firebase."}
\end{flushleft}
}}
\captionof{Quote}{An example of an app developer encountering database read/write permission issues
\cite{rq2-database}.\label{Quote:rq2-database}}
\end{center}

\noindent\textit{\textbf{App Distribution Service.}} These are services, including the Google Play Store and Apple App Store, that play a crucial role in giving app developers a platform to publish and make their apps available for end-users to discover and download. However, to protect end-user data and privacy from malicious, fraudulent, and unsafe apps (i.e., apps with security vulnerabilities), most app stores enforce security policies that apps must adhere to before they are made available to end-users.  

Examination of questions shows that developers seek assistance resolving rejections and warnings from app stores like the Google Play Store. These warnings are due to developers having security vulnerabilities in their apps (e.g., JavaScript Interface Injection Vulnerability). Most vulnerabilities are associated with third-party libraries the app uses, but there are also instances of developers incorrectly using the APIs of the mobile operating system. Developers are unsure as to how to fix these vulnerability issues and usually include a code snippet, log trace, and build script with their question, an example of which is provided in Quote \ref{Quote:rq2-distribution}.

\begin{center}
\fbox{\parbox{\dimexpr\linewidth-2\fboxsep-2\fboxrule\relax}{\centering
\textbf{Google Play Security Alert - Your app is using an unsafe implementation of the HostnameVerifier}
\begin{flushleft}
\textit{``Recently one of my app got a security alert from Google Play: `You app is using an unsafe implementation of the HostnameVerifier'. And refer a link to Google Play Help Center article for details regarding to fixing and deadline of vulnerability. Below is my code. Anyone can explain with example what changes should I do to fix this warning?"\\}
\footnotesize
HttpsURLConnection.setDefaultHostnameVerifier(new HostnameVerifier()\{ \\
    \hspace{1 cm} public boolean verify(String arg0, SSLSession arg1) \{ \\
        \hspace{2 cm} return true;\}\}); 

\end{flushleft}
}}
\captionof{Quote}{An example of a developer seeking help in resolving a Google Play Store security policy warning for their app 
(\cite{rq2-distribution}).\label{Quote:rq2-distribution}}
\end{center}

\noindent\textit{\textbf{Encryption.}} While a database provides a medium for an app to store its data, it is highly recommended to encrypt sensitive information to ensure it can only be accessed by authorized entities (\cite{Online:M9Insecu30}). Furthermore, multiple encryption algorithms are available for developers to utilize in their apps to secure data transmitted over networks or stored on the device.

While our analysis of the questions reveals developers struggling with data encryption and decryption (e.g., Quote \ref{Quote:rq2-encryption}), we also observe that a common pain point for Android and iOS app developers is the generation and storage of keys (usually private keys). Further, app developers need help using various encryption algorithms (e.g., RAS, AES, and DES) to secure data. We also observe Android developers needing help retrieving keys from the KeyStore.

\begin{center}
\fbox{\parbox{\dimexpr\linewidth-2\fboxsep-2\fboxrule\relax}{\centering
\textbf{How to encrypt and decrypt file in Android?}
\begin{flushleft}
\textit{``I want to encrypt file and store it in SD card. I want to decrypt that encrypted file and store it in SD card again. I have tried to encrypt file by opening it as file stream and encrypt it but it is not working. I want idea how to do this."}
\end{flushleft}
}}
\captionof{Quote}{An example of an Android app developer requesting help with data encryption and decryption 
(\cite{rq2-encryption}).\label{Quote:rq2-encryption}}
\end{center}

\noindent\textit{\textbf{Permission.}} To further protect user data and privacy, mobile operating systems require apps to specify the type of permissions they require to access certain information and resources. Failure to do so results in the app being denied access to these features, limiting its functionality and possibly causing a crash.  

Our analysis shows that most developers need help resolving permission-related exceptions (e.g., \texttt{java.lang.SecurityException}) their apps throw when end-users interact with it. We observe most exceptions occurring due to developers not being aware they need to request specific permissions for the app's action. We also observe developers facing issues due to bad configurations, such as missing or incorrect entries in the AndroidManifest file and needing help defining custom permissions for Android apps. For example, in Quote \ref{Quote:rq2-permission}, the developer is running into an exception due to a missing permission.

\begin{center}
\fbox{\parbox{\dimexpr\linewidth-2\fboxsep-2\fboxrule\relax}{\centering
\textbf{Android Security Exception while accessing contacts}
\begin{flushleft}
\textit{``This is totally weird and I've searched through the forums. In my main class I have a button onClick will launch the contacts application as shown below. When I click the button, the contacts list is shown but as soon as I tap on a contact a security exception is thrown. I've checked the manifest and have tried all combinations of placing the uses-permission tag, within the application, activity etc. But nothing works. Any help will be greatly appreciated."}
\end{flushleft}
}}
\captionof{Quote}{An example of a developer running into a permission exception due to missing permission 
(\cite{rq2-permission}).\label{Quote:rq2-permission}}
\end{center}

\noindent\textit{\textbf{File-Specific.}} Mobile app development is a complex process that involves the use of various file types, such as resources, build, data, certificates, and downloaded files, among others. While creating an app, developers need to integrate and manage these files effectively. However, it can be challenging for developers to understand how to integrate and manage these files. 

Some issues we observe include help with resolving zip path traversal vulnerabilities, resolving security/permission exceptions when downloading files, making configurations to correct or improve the app's security policy, and securing files. For example, in Quote \ref{Quote:rq2-file}, the developer needs help understanding how to secure the app's resource files. Finally, even though some of the challenges in this topic are also related to other topics (e.g., encryption and permissions), the association of files with app security is still a substantial challenge that deserves a topic of its own.  

\begin{center}
\fbox{\parbox{\dimexpr\linewidth-2\fboxsep-2\fboxrule\relax}{\centering
\textbf{A 3rd party application reads my asset folder}
\begin{flushleft}
\textit{``A 3rd party security application reads into my application. They probably read my asset folder. How is this possible? I thought that the sandbox model prevented from external access to the internal data structure of an app?"}
\end{flushleft}
}}
\captionof{Quote}{An example of a developer needing help with protecting/securing files that are part of the app 
(\cite{rq2-file}).\label{Quote:rq2-file}}
\end{center}

\noindent\textit{\textbf{General Security.}} While the other topics mentioned above are related to specific areas of mobile app security, these are not the only security concerns mobile app developers face. In this topic cluster, we encounter security challenges that do not occur frequently enough to be represented in separate topics. This topic includes a number of questions on authentication and access control, such as storing credentials and incorporating OAuth-based authentication with sites like Facebook and Google. We also see questions on validating and sanitizing user input to prevent client-side injections, protecting artifacts owned or utilized by the app on the device, and working with the lock screen of the device.

Further, we also see developers seeking help to obfuscate the app's source code to protect hardcoded sensitive data, such as passwords. Additionally, we observe questions on the app's distribution package file (i.e., apk and ipa), such as signature generation and verification, protecting from piracy, and preventing reverse engineering, as shown in Quote \ref{Quote:rq2-general}.

\begin{center}
\fbox{\parbox{\dimexpr\linewidth-2\fboxsep-2\fboxrule\relax}{\centering
\textbf{How to avoid reverse engineering of an APK file}
\begin{flushleft}
\textit{``I am developing a payment processing app for Android, and I want to prevent a hacker from accessing any resources, assets or source code from the APK file. Now my questions are: How can I completely prevent reverse engineering of an Android APK? Is this possible? How can I protect all the app's resources, assets and source code so that hackers can't hack the APK file in any way? Is there a way to make hacking more tough or even impossible? What more can I do to protect the source code in my APK file?"}
\end{flushleft}
}}
\captionof{Quote}{An example of a developer seeking help with securing the app's distribution package file
(\cite{rq2-general}).\label{Quote:rq2-general}}
\end{center}

\begin{tcolorbox}[top=0.5pt,bottom=0.5pt,left=1pt,right=1pt]
\textbf{Summary for RQ2.}
When it comes to mobile security, our findings show that developers turn to Stack Overflow for assistance for specific reasons, such as Secured Communications, Database, App Distribution Service, Encryption, Permissions, and File-Specific. Additionally, developers need help with other General Security practices/concerns, including code obfuscation and securing app distribution packages.
\end{tcolorbox}

\section{Threats To Validity}
\label{sec:threats}
While there are other online question-and-answer sites and forums, we limit our analysis to Stack Overflow. \rev{While extensive, it may not fully represent the broader landscape of mobile app development and security, which could impact the generalizability of our results.}
Further, our data extraction query is limited to Android and iOS questions and includes the term `security' in the title and tag. While these two mobile operating systems make up the largest market share, there can be instances of security issues specific to other mobile operating systems that we did not capture in our study. Our analysis is also constrained to analyzing only the most recent version of a question, as is the case with similar studies. In RQs that involve a manual review of data, we utilize a statistically significant random sample. Even so, our review sample may not contain important data items. 
Finally, our study presents a snapshot of current mobile app security topics and trends, providing future studies with a platform for examining the evolution of mobile app security.

\section{Conclusion \& Future Work}
\label{sec:conclusion}
This exploratory study confirms and extends findings from previous research, showing that Stack Overflow is a popular platform for mobile app developers to seek help with various security challenges. Our findings align with the study by \textcite{Yang2016security} on general software security issues, identifying similar topics such as injection vulnerabilities, SSL, certificates, databases, encryption, credentials, and access control. We also observe a sudden spike in questions related to API changes, as reported by \textcite{Linares2014ICPC}. Our results confirm that permissions are a common issue for app developers, as shown by \textcite{Scoccia2019SCAM}. Furthermore, our study aligns with \textcite{Rahkema2022iOS} findings on iOS app developers using vulnerable third-party libraries. \rev{This study contributes to the theory of knowledge sharing in software development communities by revealing how mobile app security knowledge is collaboratively constructed and disseminated on Stack Overflow.}

The findings of this study highlight practical challenges developers face in securing their apps, providing educators, researchers, and project teams with insights into specific mobile security focus areas. These insights can be used to improve training materials, courses, and code quality tools, as well as to help developers and project teams plan for problematic areas in securing their apps. \rev{Given the increasing use of Generative AI technologies, such as Large Language Models (LLMs), our future work will explore the extent to which developers rely on LLMs instead of traditional resources like Stack Overflow for help with app security.}

\printbibliography

\end{document}

%% file: hicss-packages.tex
\usepackage[utf8]{inputenc}
\usepackage{newunicodechar,graphicx}
\DeclareRobustCommand{\okina}{%
  \raisebox{\dimexpr\fontcharht\font`A-\height}{%
    \scalebox{0.8}{`}%
  }%
}
\newunicodechar{ʻ}{\okina}
\usepackage{enumitem}
\setlist[itemize,enumerate]{noitemsep, topsep=0pt, leftmargin=1.0em}

\usepackage[letterpaper]{geometry}
\usepackage{hicss}
\usepackage{times}
\usepackage[none]{hyphenat}
\usepackage{url}
\usepackage{latexsym}
\usepackage{indentfirst}
\usepackage{graphicx}
\usepackage{soul}
\usepackage{booktabs}
\usepackage{multirow}
\usepackage{float}
\usepackage{tcolorbox}
\usepackage{tabularx} 
\usepackage{balance}
\usepackage{caption}
\usepackage[hidelinks,bookmarks=true]{hyperref}
\usepackage[style=apa]{biblatex}
\usepackage{varwidth}
\newsavebox\tmpbox
\DeclareCaptionType{Quote}
\usepackage{background}
\backgroundsetup{
  position=current page.east,
  angle=-90,
  nodeanchor=east,
  vshift=-4mm,
  opacity=0.5,
  scale=3,
  contents=PREPRINT
}